\documentclass{cys}

\usepackage[utf8]{inputenc}

\usepackage[english]{babel}

\usepackage{graphicx}

\usepackage{epstopdf} 

\addto\captionsenglish{%
}

\addto\captionsspanish{%
}

\usepackage{url}
\usepackage{amsmath}

\title{Lightweight Online Separation of the Sound Source of Interest through BLSTM-Based Binary Masking}

\author{Alejandro~Maldonado$^1$, Caleb~Rascon$^2$ and Ivette~Velez$^1$}

\affil{ 
	$^1$ Posgrado en Ciencia e Ingeniera de la Computacion,\authorcr
	Universidad Nacional Autonoma de Mexico, Ciudad de Mexico, Mexico
	\authorcr \authorcr
	$^2$ Instituto de Investigaciones en Matematicas Avanzadas y en Sistemas,\authorcr
	Universidad Nacional Autonoma de Mexico, Ciudad de Mexico, Mexico
	\authorcr \authorcr
	caleb.rascon@iimas.unam.mx
	\authorcr  \authorcr
}

\begin{document}

\maketitle

\renewcommand{\tablename}{Table}

\begin{abstract}
    Online audio source separation has been an important part of auditory scene analysis and robot audition. The main type of technique to carry this out, because of its online capabilities, has been spatial filtering (or beamforming), where it is assumed that the location (mainly, the direction of arrival; DOA) of the source of interest (SOI) is known. However, these techniques suffer from considerable interference leakage in the final result. In this paper, we propose a two step technique: 1) a phase-based beamformer that provides, in addition to the estimation of the SOI, an estimation of the cumulative environmental interference; and 2) a BLSTM-based TF binary masking stage that calculates a binary mask that aims to separate the SOI from the cumulative environmental interference. In our tests, this technique provides a signal-to-interference ratio (SIR) above 20 dB with simulated data. Because of the nature of the beamformer outputs, the label permutation problem is handled from the beginning. This makes the proposed solution a lightweight alternative that requires considerably less computational resources (almost an order of magnitude) compared to current deep-learning based techniques, while providing a comparable SIR performance.
\end{abstract}


\section{Introduction}

Sound source separation is an essential step in the processing chain of events in computational auditory scene analysis~\cite{wang2005ideal} (CASA) and robot audition~\cite{1389723,4960426} (RA). Currently, many sound-related tasks such as automatic speech recognition, speaker identification, and mood classification, assume that their input bares only the audio data from the source to be analyzed. Many of the techniques used for carrying these tasks are based on machine learning methods, which could be made robust against multiple-source scenarios by augmenting their corresponding training corpora. However, another alternative could be to have a source separation phase beforehand that provides the audio information from one source at a time. To this effect, current techniques could still be used by this alternative, without requiring impractical amounts of space and time for training.

In terms of sound source separation, it is of interest to carry it out in an \textit{online} manner (meaning, ``on the fly''), for scenarios in which the user is interacting with a CASA/RA system, such as a service robot, a virtual assistant, a security system, etc. This is opposed to an \textit{offline} manner, which records the audio from the environment and returns the results of the audio analysis after the interaction is completed. It is important to note that we are differentiating between carrying this analysis out in an \textit{online} manner and carrying it out in \textit{real-time}, since the latter involves discussion of specific response times thresholds~\cite{8172394}. What we define as \textit{online} analysis is that the system's response time is less than the length of the time window that is to be processed. Meaning that, even though results are given while the user is interacting, they may be given with a certain delay. In certain applications, like human-robot interaction (HRI), online results are essential for the interaction to be successful, while a reasonable delay ($\leq 1$ second) is acceptable~\cite{8172394}.

One of the most popular type of techniques to carry out online source separation is by spatial filtering or \textit{beamforming}. It is important to note that the terms ``source separation'' and ``beamforming'' are not usually used in the same context. While source separation aims to separate all sources present in the recorded environment, beamforming aims to separate one source of interest (the location of which is known \textit{a priori}) from the rest of the environment. In this paper, we are equating these two terms, since beamforming is carrying out a type of source separation, and it is usually designed to be carried out in an online manner. And, while beamforming only separates one sound source from the environment, it is compatible with the aforementioned CASA/RA/HRI application scenarios in which one user is attended at a time, i.e. the source of interest (SOI).

Unfortunately, an important issue with beamforming is that of interference leakage, in which sound sources different from the SOI are still present in the final result. Although this interference presence can be low if the beamformer is configured appropriately~\cite{7372431}, it is still high enough to be perceivable (with signal-to-interference ratios less than 15 dB), which may have an impact in subsequent CASA/RA modules~\cite{7179045}. Thus, beamforming techniques tend to employ a high number of microphones to overcome this issue~\cite{RASCON2017184}.

On the other hand, deep learning strategies have shown impressive results when carrying out source separation, even when only a single microphone is used~\cite{6853860,7471631,7952155}. A popular methodology is to classify which frequency bins belong to which sound source, i.e. frequency masking. To carry out this through time, many of these techniques track the frequential variation of each source, so that in each time window the appropriate time-frequency (TF) bins is assigned to the correct source. If this tracking is done incorrectly, one source may be assigned data from others in different time windows, corrupting the overall output. Solving this requires complex solutions that require an important amount of computational resources (which may be an important issue for some CASA/RA/HRI application scenarios). Or, in the worst case, the problem is bounded such that the proposed techniques are only tested with recordings with a few amount of interferences~\cite{6853860,7471631,7952155}.

To overcome this problem, we proposed a novel lightweight source separation technique which carries out deep-learning-based frequency masking from a beamforming output. The proposed solution can be run online and is robust against variations of interferences and number of microphones. It is composed of two parts: 1) a phase-based time-frequency-masking beamforming that provides both the estimation of the SOI and the estimation of the cumulative environmental interference; and 2) a time-frequency binary masking stage based on a bidirectional long short-term memory (BLSTM) network, that aims to use these two estimations to separate the SOI from the environmental interference estimation. Since the proposed beamformer is already providing a preliminary separation of the TF bins of the SOI from the TF bins of the interferences, the permutation problem~\cite{7471631} is solved from the beginning. This means that the complexity and size of the BLSTM network architecture is low enough to be run in an online manner, even with modest computer equipment.

The full system is found in \url{http://github.com/balkce/onlinessblstm}.

The work here presented has the following structure: Section \ref{sec:background} provides a brief summary of the related works and background relevant to the proposed technique; Section \ref{sec:system} details the proposed technique; Section \ref{sec:eval} describes the evaluation methodology against a deep-clustering-based source separation approach and presents the results; Section \ref{sec:discuss} discusses the insights obtained from these results; and Section \ref{sec:conclusion} provides our conclusions and future work.

\section{Background and Related Work}
\label{sec:background}

As mentioned before, we are aiming to use a beamforming paradigm to carry out online sound source separation, which implies that the location of the source of interest (SOI) is known \textit{a priori}. This approach is popular in Robot Audition (RA), as shown by HARK~\cite{Nakadai_2017jrm} and ManyEars~\cite{Grondin2013}, both of which employ a real-time variation of the geometric source separation technique~\cite{4285864}. This technique merges both the beamforming paradigm with a blind sound source separation approach. It is worth mentioning that HARK has modified this technique even further by introducing adaptability to the inner mechanisms of geometric source separation~\cite{5299088}, and that ManyEars has pushed for being more lightweight, with its ODAS project~\cite{GRONDIN201963}\footnote{In this reference, ODAS does not report any source separation capabilities, but its authors have already added this functionality to its base code~\cite{odasgithub}.}. In all these circumstances, the direction of arrival (DOA) of the sound source is assumed to be known \textit{a priori}, or estimated by applying one of the many sound source localization techniques reported in literature~\cite{RASCON2017184}. However, evaluation of these beamforming techniques has been bounded by the use of a considerable amount of microphones, which reduces the presence of interferences in the resulting SOI estimation.

It would be of interest to use less microphones, while avoiding the aforementioned interference leakage issue, when carrying out online sound source separation. A possible alternative to this would be to apply recent developments in mono-aural sound source separation, the vast majority of which employ deep-learning techniques such as bidirectional long short-term memory (BLSTM) networks~\cite{1556215}. This type of techniques are a type of recurrent network which are ideal to be used with temporal data, and have been a good answer to issues specific to recurrent networks, such as the vanishing gradient problem~\cite{5264952} and localized classification~\cite{Graves:2006:CTC:1143844.1143891}. To this effect, they have shown very good results for speech recognition~\cite{6638947,6707742} and text recognition~\cite{7050699}. However when applied to sound source separation, an important issue has been found: the permutation problem~\cite{8369155, 7471631}. Many of these techniques aim to classify each time-frequency (TF) bin of the input signal as to belonging to one of various possible sources. When carrying out this process throughout several time windows, it is essential that the TF bins are appropriately assigned such that classifications of previous time windows are correctly followed; if not, one source may be assigned TF bins of other sources in subsequent time windows, with unwanted overall results. To avoid this, several methods have been used such as permutation invariant training~\cite{7952154}, deep clustering~\cite{7471631} and deep attractors~\cite{7952155}.

Specifically, deep clustering has been successful in recent years for sound source separation. The Chimera network~\cite{7952118} is a representative example of this. Even though it was originally proposed for voice separation in music, it has been recently modified for its use in mono-aural source separation~\cite{8462507,8540037,8683231}. The Chimera network uses a two-front approach to design its objective loss function: 1) a type of deep-clustering loss function that transforms the input signal to a domain in which it is able to keep track of which sound source is which (by means of clustering methodologies), and 2) a magnitude spectrum approximation objective that aims to infer the TF mask to apply to the input signal. By training with this loss function, the network is made to consider which source is being assigned to which TF bin, resulting in strong Signal-to-Distortion performances.

However, as it can be deduced, an important amount of complexity is encountered when carrying out this methodology. This results in a complex solution space that the training optimization algorithm is expected to solve. It would be of interest to avoid this issue altogether, which not only would simplify the solution space to solve, but also may reduce its memory requirements and its response time.

It is important to mention the work of~\cite{PERTILA201597}, which carries out a similar technique to ours. However, the authors of this work feed the network with features extracted from the beamformer weights. Although this process solves the permutation problem from the beginning, it implicitly trains the network with the array geometry (since the beamformer weights are based on it). It would be of interest to solve the permutation problem while the system is robust against array geometry changes.

Additionally, a hybrid approach of a beamformer and deep learning techniques has been employed before~\cite{8369155}. However, this hybrid approach is usually carried out by either: 1) making the deep learning network emulate the task of the beamformer, or 2) feeding the beamformer estimation of the source of interest as the mono-aural input to the deep learning network. As far as we know, feeding the deep learning network a two-channel input (one of the preliminary estimation of the SOI, and the other of the preliminary estimation of the cumulative environmental interference) has not been proposed before.

\section{Proposed System}
\label{sec:system}

\begin{figure}[ht]
\centering
\includegraphics[width=0.4\textwidth]{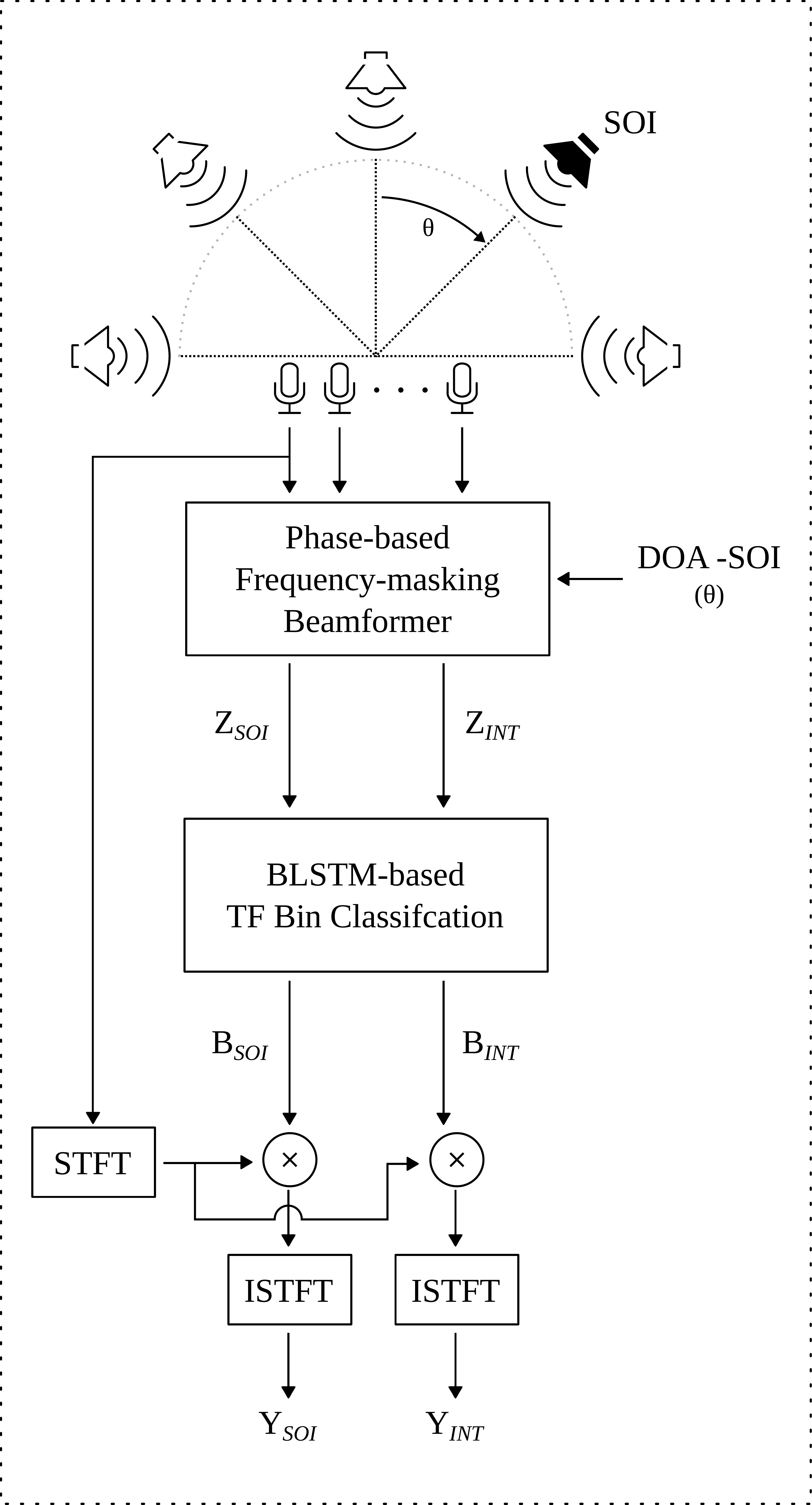}
\caption{An overall diagram of the proposed system.}
\label{fig:proposedsystem}
\end{figure}

An overall summary of the proposed system is shown in Figure \ref{fig:proposedsystem}. As it can be seen, there are two core modules. First, the audio data from the microphone array and the direction of arrival of the source of interest (SOI) is fed to a phase-based frequency-masking beamformer that provides a preliminary estimation of the SOI ($\mathbf{Z}_{SOI}$) as well as of the cumulative environmental interference ($\mathbf{Z}_{INT}$). Second, a time-frequency binary masking stage, based in a bidirectional long short-term memory (BLSTM) network, provides a time-frequency (TF) binary mask ($\mathbf{B}_{SOI}$) that separates the SOI from the cumulative environmental interference estimation. This mask is then applied to the signal of the reference microphone for the final SOI estimation ($\mathbf{Y}_{SOI}$). In this section, these two core modules are detailed.

\subsection{Phase-Based Frequency Masking Beamformer}
\label{sec:beamformer}

\begin{figure}[ht]
\centering
\includegraphics[width=0.4\textwidth]{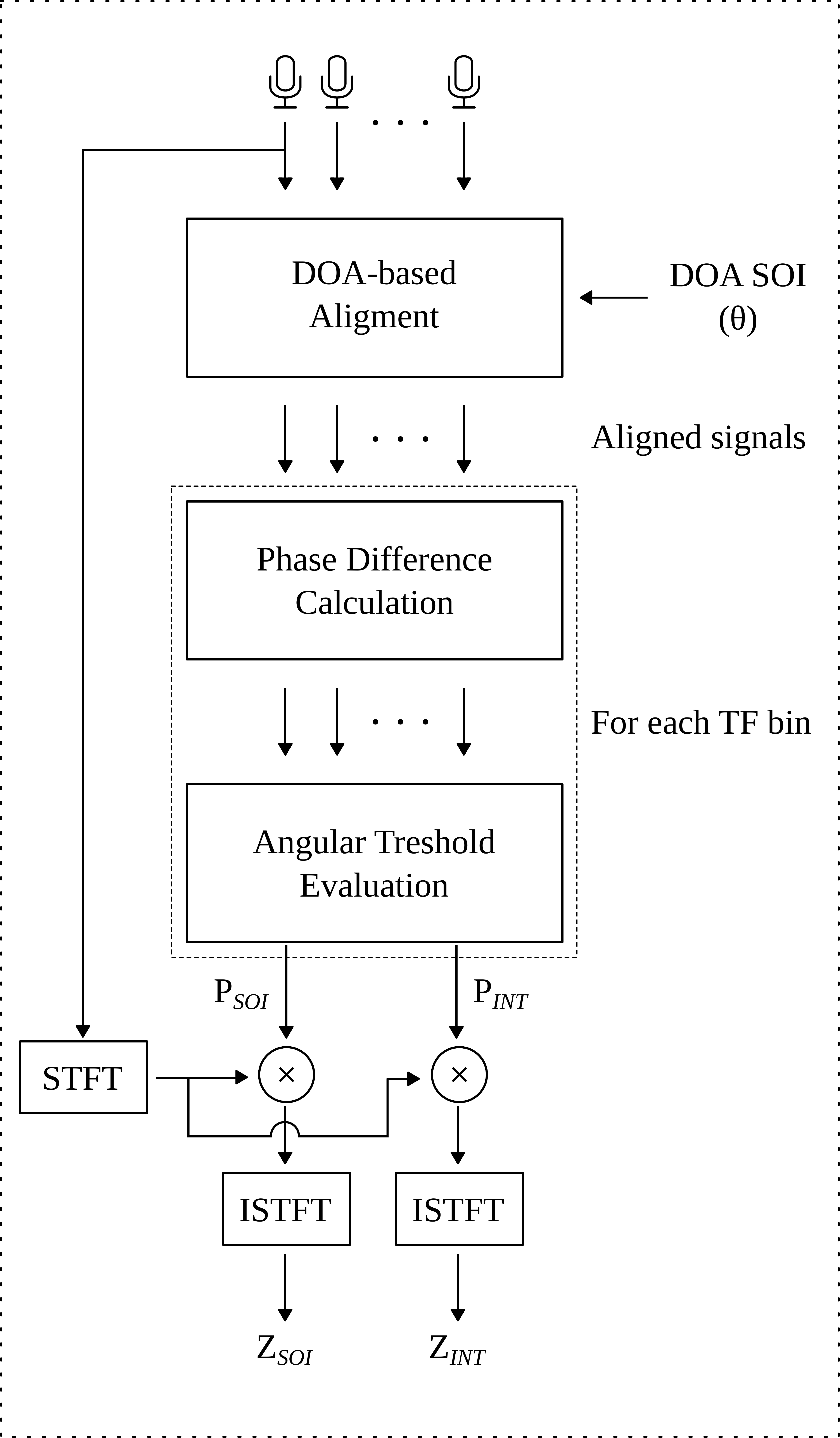}
\caption{A diagram summarizing the phase-based frequency-masking beamformer stage.}
\label{fig:beamformer}
\end{figure}

The proposed beamformer is summarized in Figure \ref{fig:beamformer}.

Let $\mathbf{X}$ be the input matrix of size $M \times N$, where $M$ is the number of microphones and $N$ is the time-window length in samples, as well as the length of the resulting frequency masks. The columns of $\mathbf{X}$ are the Fourier transformed time-windows of each microphone input. Additionally, let $\theta$ be the direction of arrival (DOA) of the SOI. The first stage of the beamformer carries out a time-alignment of the columns of $\mathbf{X}$ such that the information received by the microphone array in the planar direction $\theta$ is in phase. This is carried out as described in Equation \ref{eq:alignbyphase}.

\begin{equation}\label{eq:alignbyphase}
    \mathbf{X}_a[m;f] = \mathbf{X}[m;f]e^{i 2 \pi f t_{m;\theta}}.
\end{equation} 

Where $\mathbf{X}_a$ is the phase-aligned version of $\mathbf{X}$ towards $\theta$, $m$ is the microphone index, $f$ is the frequency bin, and $t_{m;\theta}$ is the delay in seconds applied to the input data of microphone $m$ based on $\theta$. Using the positions of the microphones relative to the reference microphone, each respective delay can be calculated via different methodologies, such as the far-field model\cite{RASCON2017184} presented in Equation \ref{eq:farfield}.

\begin{equation}\label{eq:farfield}
    t_{m;\theta} = -\frac{r_m}{c}cos(\theta_m-\theta).
\end{equation} 

Where $c$ is the speed of sound ($\sim$ 343 meters per second), and $r_m$ and $\theta_m$ are the polar coordinates of microphone $m$ in relation to the reference microphone ($m=1$).

The average phase difference is then calculated for each frequency bin $f$, as described in Equation \ref{eq:phasediff}.

\begin{equation}\label{eq:phasediff}
    |\varphi|_f = \frac{2}{M(M-1)} \sum_{i=1}^{M-1}\sum_{j=i+1}^{M}|\varphi_{i;f}-\varphi_{j;f}|.
\end{equation} 

Where $M$ is the number of microphones, $|\varphi|_f$ is the average phase difference at frequency bin $f$, and $\varphi_{m;f}$ is the phase at frequency bin $f$ of microphone $m$.

Consequently, two frequency masks are created via an angular threshold ($\varphi_{max}$), as described in Equations \ref{eq:masksoi} and \ref{eq:maskint}.

\begin{equation}\label{eq:masksoi}
    \mathbf{P}_{SOI}[f]=\left\{
        \begin{matrix}
            1, & if |\varphi|_f \leq \varphi_{max}  \\
            0, & otherwise,
        \end{matrix}\right.
\end{equation} 

\begin{equation}\label{eq:maskint}
    \mathbf{P}_{INT}[f]=\left\{
        \begin{matrix}
            0, & if |\varphi|_f \leq \varphi_{max}  \\
            1, & otherwise,
        \end{matrix}\right.
\end{equation} 

Where $\mathbf{P}_{SOI}$ and $\mathbf{P}_{INT}$ are the $1 \times N$ frequency masks for the SOI and for the cumulative environmental interference, respectively.

The $1 \times N$ estimations of the SOI ($\mathbf{Z}_{SOI}$) and the cumulative environmental interference ($\mathbf{Z}_{INT}$) are calculated by applying the corresponding frequency mask to the reference microphone, as described in Equations \ref{eq:beamsoi} and \ref{eq:beamint}.

\begin{equation}\label{eq:beamsoi}
    \mathbf{Z}_{SOI}[f] = \mathbf{P}_{SOI}[f]*\mathbf{X}[1;f].
\end{equation} 

\begin{equation}\label{eq:beamint}
    \mathbf{Z}_{INT}[f] = \mathbf{P}_{INT}[f]*\mathbf{X}[1;f].
\end{equation} 

Variations of this beamformer have been proposed before. The authors of~\cite{Brutti2016} use a similar method, but instead of creating a binary mask, they create a soft mask by assuming a frequency-dependent phase variance and empirically accounting for it. It is important to note, however, that this work does not provide an estimation of the cumulative environmental interference. Another similar work is that of~\cite{8784809}, where the authors employ an interference-leakage removal strategy that requires the estimation of the frequency co-variance matrix. This is similar to the strategy employed by the well known minimum variance distortionless response (MVDR) beamformer~\cite{levin1964maximum,1449208}, which has been shown to be too complex to be run in an online manner using the whole frequency spectrum~\cite{mvdrnotonline}. It is important to note that variations of MVDR have been developed to run online, but the strategy employed in~\cite{8784809} has not been shown to do so.

As it can be concluded, the phase-based frequency masking beamformer proposed here is much less complex than those presented in the aforementioned works. Additionally, and more importantly, it provides the estimation of the cumulative environmental interference. As it will be discussed later, this is essential to solve the permutation problem for the BLSTM-based TF binary-masking stage, resulting in having a relatively low complexity.

It is important to mention that, although $X$ represents a time-window length of $N$ samples of input data, the input length $N_{B}$ used for the subsequent binary masking stage is conformed of several of these $N$-length windows, using a Hann-window-based overlap-and-add strategy (to avoid discontinuities when applying the short-time Fourier transform). To this effect, $N_{B}$ can be considered independent of $N$, in only that $N_{B}$ is a multiple of $N$. 

\subsection{BLSTM-Based TF Binary Masking}

\begin{figure}[ht]
\centering
\includegraphics[width=0.28\textwidth]{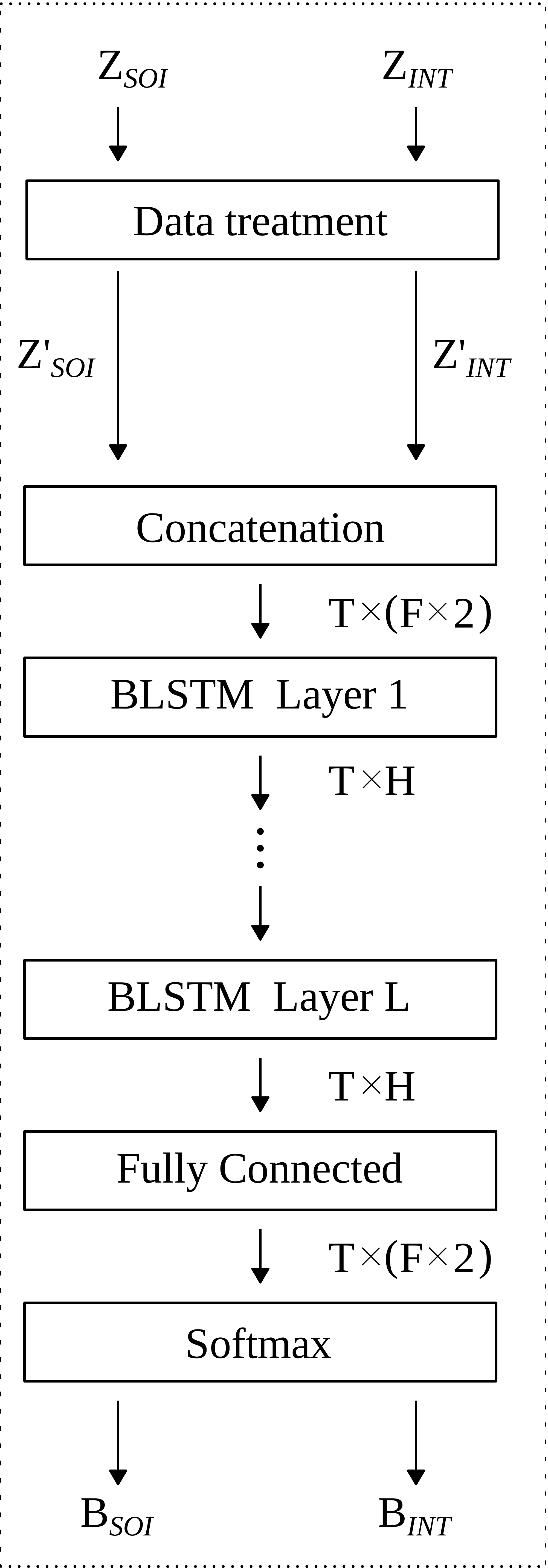}
\caption{Architecture of the proposed BLSTM network.}
\label{fig:blstm}
\end{figure}

In Figure \ref{fig:blstm}, the BLSTM-based time-frequency binary masking stage is summarized. In an overall sense, the purpose of this stage is to calculate a time-frequency binary mask ($\mathbf{B}_{SOI}$) which, when applied to the input data of the reference microphone, the SOI is separated from the cumulative environmental interference.

As it can be seen in Figure \ref{fig:blstm}, the BLSTM-based TF binary masking stage expects two inputs, one with the SOI estimation and another with the estimation of the cumulative environmental interference. These two time-domain inputs of length $N_{B}$ are transformed to the time-frequency (TF) domain via the short-time Fourier transform, using a Hann window with a length of $N_H$ samples and a 50\% overlap. This results in two matrices of size $T \times F$. The size of time dimension $T$ depends on $N_{B}$ such that $T = (N_{B}*2)+1$, because of the 50\% overlap (with zero-padded Hann windows at the edges). As for the size of the frequency dimension $F$, it depends on $N_H$: to avoid redundant weight calculations in the subsequent BLSTM network, only the lower half (with the DC component) of the mirrored Fourier transform is used, thus $F = \frac{N_H}{2}+1$.

The energy at each input TF bin is converted into the decibel scale (dB), and then standardized to have zero-median and one-standard-deviation. These two steps are important to mold the solution space to a shape that is easier to converge when training the subsequent BLSTM network. The standardized inputs are then concatenated in the frequency dimension.

The proposed BLSTM network is made up of $L$ amount of BLSTM stacked layers with $H$ amount of hidden units, which are then fed into a fully-connected layer, and has a softmax output layer that estimates the probability of the TF bin belonging to the SOI. Thus, the BLSTM network carries out a binary classification, which results in two $T \times F$ binary masks: one for the SOI ($\mathbf{B}_{SOI}$) and one for the cumulative environmental interference ($\mathbf{B}_{INT}$), although only $\mathbf{B}_{SOI}$ is used in later stages.

Once trained, as shown in Figure \ref{fig:proposedsystem}, the resulting $\mathbf{B}_{SOI}$ is applied to the input data of the reference microphone, which is transformed to the time-frequency domain in the same manner as the outputs of the beamformer. This process, as described by Equation \ref{eq:resultsoi}, results in the final SOI estimation $\mathbf{Y}_{SOI}$ in the time-frequency domain.

\begin{equation}\label{eq:resultsoi}
    \mathbf{Y}_{SOI}[t;f] = \mathbf{B}_{SOI}[t;f]*\mathbf{X}[1;t;f].
\end{equation}

If the application requires it, the final estimation of the cumulative environmental interference ($\mathbf{Y}_{INT}$) can be obtained by applying $\mathbf{B}_{INT}$ to the reference microphone, as shown in Equation \ref{eq:resultint}.

\begin{equation}\label{eq:resultint}
    \mathbf{Y}_{INT}[t;f] = \mathbf{B}_{INT}[t;f]*\mathbf{X}[1;t;f].
\end{equation}

\subsubsection{Training and Validation}

For training, the LibriSpeech corpus~\cite{7178964} was used, which is composed of 500 hours of clean recordings of users reading text, sampled at 16 kHz. The users were chosen randomly from 80\% of this corpus to act as sound sources which were artificially mixed to simulate the inputs of a 2-microphone array; the other 20\% was used for validation purposes. For the second microphone, each source was delayed according to a randomly chosen DOA for each sound source, applying the far-field model shown in Equation \ref{eq:farfield}. The DOA was chosen in the $[-90^o,90^o]$ range, at $45^o$ intervals in the horizontal plane.

Additionally, the ideal TF mask ($\mathbf{O} _{k}$) of each source $k$ was calculated from the clean corpus recordings, and used as part of a magnitude spectrum approximation (MSA) objective function ($\mathcal{L}$), described in Equation \ref{eq:loss_function}.

\begin{equation}\label{eq:loss_function}
    \mathcal{L} = \sum_{k=1}^2 || (\mathbf{O} _{k} - \mathbf{B}_{k}) \odot \mathbf{S} ||_2^2.
\end{equation}

Where $k$ is either 1 for $SOI$ or 2 for $INT$; $\mathbf{B} _{k}$ is the predicted mask; and $\mathbf{S}$ indicates the magnitude of the TF bin of the mixture from the reference microphone. This is similar as to what was carried out in~\cite{7952118}.

During training, before delivering $\mathbf{B}_{SOI}$ to the loss function, a simple voice-activity detection (VAD) mechanism~\cite{7471631} is employed, described in Equation \ref{eq:vad}.

\begin{equation}\label{eq:vad}
	\begin{aligned}
    &\psi[t;f]=\left\{
        \begin{matrix}
            1, & if~||\mathbf{X}[1;t;f]|| > \mathbf{X}[1]_{max} - V  \\
            0, & otherwise,
        \end{matrix}\right.\\
     &\mathbf{B}_{SOI}[t;f]=\psi[t;f]\mathbf{B}_{SOI}[t;f] 
     \end{aligned}
\end{equation} 

Where $\psi[t;f]$ is the VAD mask, the operator $||\cdot||$ calculates the decibel energy of a TF bin, $V$ is the VAD energy threshold, and $\mathbf{X}[1]_{max}$ is the maximum decibel energy of the reference microphone $\mathbf{X}[1]$ in an input length.

It is important to mention that the VAD step is only necessary during training, and not during testing. This is because, given the design of the loss function, the BLSTM network implicitly learns to ignore the TF bins usually discarded by the VAD process.

During training, the RMSProp optimizer was used with a learning rate of $10\mathrm{e}^{-5}$ and a momentum of $0.9$, as employed by~\cite{7471631}.

\subsubsection{Architecture Selection}
\label{sec:archselec}

To select the architecture for the proposed BLSTM network, we evaluated different architecture configurations, trained with up to 3 sources (including the source of interest, meaning, with up to 2 interferences). In Table \ref{architecture_results_3}, their performance is reported in terms of the signal-to-interference ratio (SIR) in the output. This was measured using the BSS\_EVAL\_SOURCES algorithm~\cite{1643671} using the clean recordings of LibriSpeech as the basis of comparison. This table also reports the memory\footnote{We define ``memory'' as the amount of RAM (measured in MB) the model occupies when not carrying out any operations, as a representation of the computational resources it requires to run.} occupied by each model.

\begin{table}[ht]
\renewcommand{\arraystretch}{1.3}
    \centering
    \caption{Evaluation of different configurations of proposed BLSTM model with up to \textbf{3} sources.}
    \begin{tabular}{ccccc}
        \hline
        $\mathbf N_{B}$ & $\mathbf H$ & $\mathbf L$ & \bf Memory (MB) & \bf SIR (dB) \\ 
        \hline
        8192 & 200 & 1 & 16 & 19.69 \\ 
        8192 & 200 & 3 & 38 & 22.44 \\ 
        8192 & 200 & 5 & 60 & 22.68 \\ 
        8192 & 300 & 1 & 26 & 20.87 \\ 
        \textit{8192} & \textit{300} & \textit{3} & \textit{76} & \textit{23.67} \\ 
        8192 & 300 & 5 & 125 & 22.06 \\ 
        8192 & 400 & 1 & 39 & 21.82 \\ 
        8192 & 400 & 3 & 127 & 22.94 \\ 
        8192 & 400 & 5 & 215 & 22.17 \\ 
        8192 & 500 & 4 & 259 & 26.02 \\ 
        16384 & 200 & 1 & 16 & 20.99 \\ 
        \textbf{16384} & \textbf{200} & \textbf{3} & \textbf{38} & \textbf{24.66} \\ 
        16384 & 200 & 5 & 60 & 22.88 \\ 
        16384 & 300 & 1 & 26 & 21.36 \\ 
        16384 & 300 & 3 & 76 & 23.38 \\ 
        16384 & 300 & 5 & 125 & 21.93 \\ 
        16384 & 400 & 1 & 39 & 22.68 \\ 
        16384 & 400 & 3 & 127 & 23.82 \\ 
        16384 & 400 & 5 & 215 & 22.65 \\ 
        16384 & 500 & 4 & 259 & 27.75 \\ 
        \hline
         \multicolumn{3}{r}{\textit{Average}} & 98.1 & 22.82 \\ 
        \hline
    \end{tabular}
    \label{architecture_results_3}
\end{table}

The configurations vary in terms of number of BLSTM stacked layers ($L$), number of hidden units ($H$) and input length ($N_{B}$). The results when varying other parameters are not reported since they did not provide considerable differences in the evaluations. Meaning, in these evaluations, the Hann-window length $N_H$ was set at 512 samples and $V$ is set at 40 dB. In~\cite{7952118} the authors employed 4 BLSTM stacked layers and 500 hidden units, and obtained robust performances in mismatched conditions. Since the aim of this work is to minimize memory usage, these were chosen as the combined upper bound for $L$ and $H$. For $H<500$, we tested $L$ values of 1, 3 and 5 to provide a balanced view of the performance fluctuation when varying $L$. We also set $\varphi_{max}$ to $60^o$.

It is of interest to select an architecture configuration that both maximizes its SIR performance while minimizing its memory usage. To this effect, we calculate the area under the curve as defined in Equation \ref{eq:lineselect} for each of the architecture configurations in Table \ref{architecture_results_3}.

\begin{equation}\label{eq:lineselect}
    y(x)=\left\{
        \begin{matrix}
            0, & if ~ x < 0  \\
            \left(\dfrac{\sigma_a}{\mu_a}\right)x, & if ~ 0 < x < \mu_a  \\
            \sigma_a, & otherwise
        \end{matrix}\right.
\end{equation} 

Where $\sigma_a$ and $\mu_a$ are (respectively) the SIR and memory usage for each architecture configuration $a$ presented in Table \ref{architecture_results_3}. The architecture configuration shown in bold in Table \ref{architecture_results_3} ($L$: 3, $H$: 200, $N_{B}$: 16384) has the largest area under the curve and, thus, the one we recommend to use. However, consideration should be given to the configuration shown in italics ($L$: 3, $H$: 300, $N_{B}$: 8192), since it not only provides the second largest area under the curve, but it also uses a smaller $N_{B}$ (which is close to $0.5$ seconds when sampling at 16 kHz).

\section{Evaluation and Results}
\label{sec:eval}

To investigate the behavior of the proposed system, three evaluations were carried out, two of which use the Chimera model~\cite{7952118} as a point of comparison, since it is arguably a representative example of current deep-learning-based sound source separation techniques~\cite{8462507,8540037,8683231}. The evaluated Chimera network is a modified version to the one originally presented in~\cite{7952118}, such that it was able to receive both outputs of the beamformer described in Section \ref{sec:beamformer}. It is important to mention that the original version of the Chimera network was not built for generalized sound source separation. However, with slight modifications, such as the one proposed in this work, as well as more complex such as the ones shown in~\cite{8462507,8540037,8683231}, its performance can be quite impressive.

To this effect, a similar evaluation to the one described in Section \ref{sec:archselec} (whose results are shown in Table \ref{architecture_results_3}) was carried out for the Chimera network, trained with up to 3 sources. Different configurations were evaluated, which varied in terms of $N_{B}$ and the embedding dimension used by one of the heads of the Chimera network ($D$). $H$ and $L$ were kept at 4 and 500, respectively, since these are the recommended values used in~\cite{7952118}.

The results of these evaluations, carried out with up to 3 sources (including the source of interest; meaning, 2 interferences), are shown in Table \ref{architecture_results_chim}.

\begin{table}[ht]
	\renewcommand{\arraystretch}{1.3}
	\centering
	\caption{Evaluation of different configurations of the modified Chimera model with up to 3 sources.}
	\begin{tabular}{ccccc}
		\hline
		\bf Srcs. & $\mathbf N_{B}$ & $\mathbf D$ & \bf Mem. (MB) & \bf SIR (dB) \\ 
		\hline
		3 & 8192 & 5 & 268 & 24.55 \\ 
		3 & 8192 & 10 & 283 & 25.59 \\ 
		3 & 8192 & 20 & 312 & 24.92 \\ 
		3 & 8192 & 40 & 371 & 25.29 \\ 
		\textbf{3} & \textbf{16384} & \textbf{5} & \textbf{268} & \textbf{27.25} \\ 
		3 & 16384 & 10 & 283 & 27.12 \\ 
		\underline{3} & \underline{16384} & \underline{20} & \underline{312} & \underline{27.29} \\ 
		3 & 16384 & 40 & 371 & 27.28 \\ 
		\hline
		\multicolumn{3}{r}{\textit{Average}} & 308.5 & 26.16 \\ 
		\hline
	\end{tabular}
	\label{architecture_results_chim}
\end{table}

In this section some perspectives are provided that show the applicability of the proposed system. The results of three evaluations are reported:

\begin{itemize}
    \item The relationship of the SIR performance against memory usage, for both Chimera and the proposed system.
    \item The relationship of the SIR performance against number of sources, for both Chimera and the proposed system.
    \item The robustness against changes in array geometry of the proposed system.
\end{itemize}

For these evaluations, 100 speakers were randomly chosen from the validation subset, and for each speaker 10 consecutive $N_{B}$-length windows were selected for the 16384-input length models, and 20 of these were chosen for the 8192-input length models. Both of these types of segments are approximately 10 seconds. When varying the number of sources, these segments were mixed with the segments of other randomly selected speakers from the validation subset.

\subsection{SIR vs Memory Usage}

\begin{figure}[ht]
\centering
\includegraphics[width=0.48\textwidth]{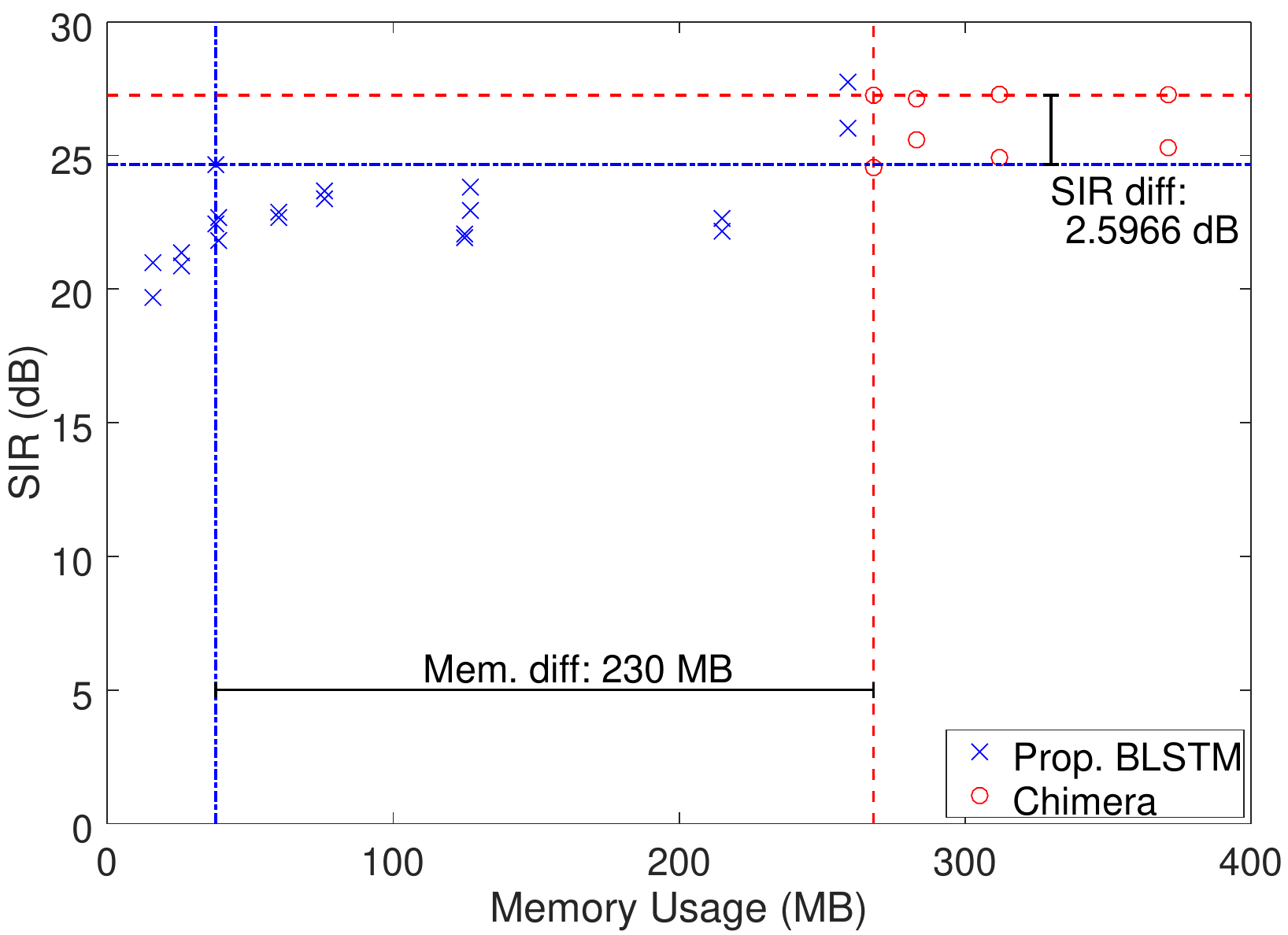}
\caption{Memory Requirements vs SIR. The blue dot-dashed lines represent the respective SIR and memory usage of the recommended configuration BLSTM-based architecture, and red dashed lines the memory usage and SIR of the similarly selected recommended Chimera architecture.}
\label{fig:SIRvsRAM}
\end{figure}

In Figure \ref{fig:SIRvsRAM} each data point represents an architecture configuration shown in Tables \ref{architecture_results_3} and \ref{architecture_results_chim}; blue crosses belong to the proposed BLSTM-based models, and red circles to the Chimera-based models. The horizontal axis represents memory usage and the vertical axis its SIR. The blue dot-dashed lines represent the memory usage and SIR of the recommended configuration of the proposed BLSTM-based architecture; and the red dashed lines represent the memory usage and SIR of the similarly selected recommended configuration from the Chimera variations (shown in bold in Table \ref{architecture_results_chim}).

As it can be seen, although the difference between the SIR of Chimera and the proposed BLSTM-based architecture configuration is low ($\sim 3$ dB), the difference between their memory usage is substantial ($> 200$ MB).

\subsection{SIR vs Number of Sources}

\begin{figure}[ht]
\centering
\includegraphics[width=0.48\textwidth]{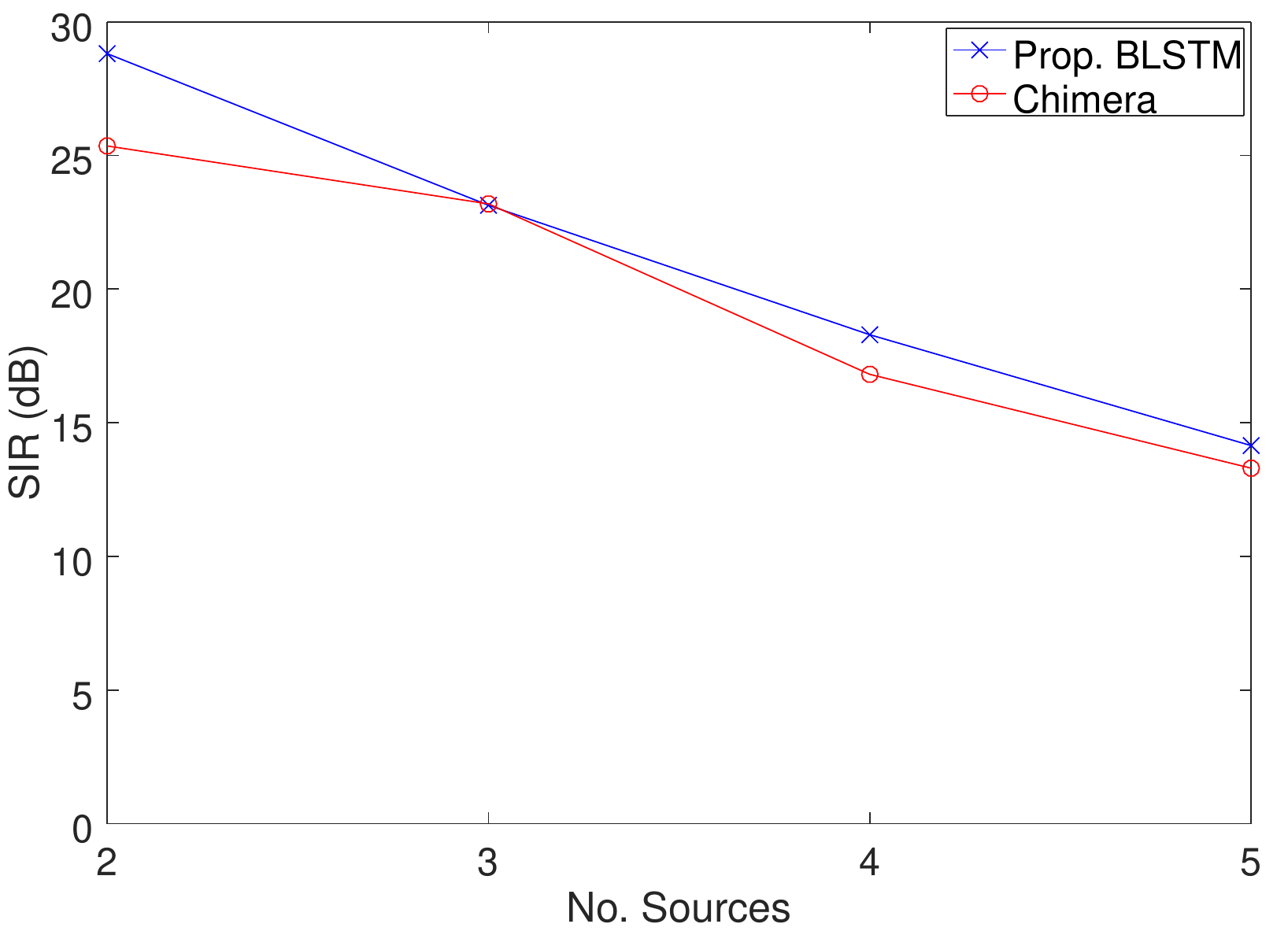}
\caption{Number of sources vs SIR of the output of the trained models.}
\label{fig:SIRvsSOURCE}
\end{figure}

It is also of interest to investigate the impact that the number of sources has on the performance. To this effect, we compare the performance of the recommended configuration of our proposed system (shown in bold in Table \ref{architecture_results_3}) as well as the best performing configuration of Chimera (underlined in Table \ref{architecture_results_chim}), as the amount of sources is increased. The results are shown in Figure \ref{fig:SIRvsSOURCE}.

It is important to note that both models were trained with up to 3 simultaneous sources, so these results reflect their ability to extrapolate the separation capabilities with more sources than they were trained with.

As it can be seen, both models have comparable SIR performance, and the obvious tendency is that as the number of sources increases, the SIR decreases (which is to be expected). An explanation for this is that the beamformer provides both an estimation of the source of interest, as well as an estimation of cumulative environmental interference from which the SOI should be separated. This means that the permutation problem is solved from the beginning. Thus, the deep clustering part of Chimera that aims to solve this problem is rendered unnecessary for this test scenario. 

\subsection{SIR vs Number of Microphones}

Since the models were trained using the output of the beamformer that was fed the simulated inputs of a two-microphone array, it is of interest to investigate the impact of the system if the number of microphones varies.

\begin{figure}[ht]
\centering
\includegraphics[width=0.48\textwidth]{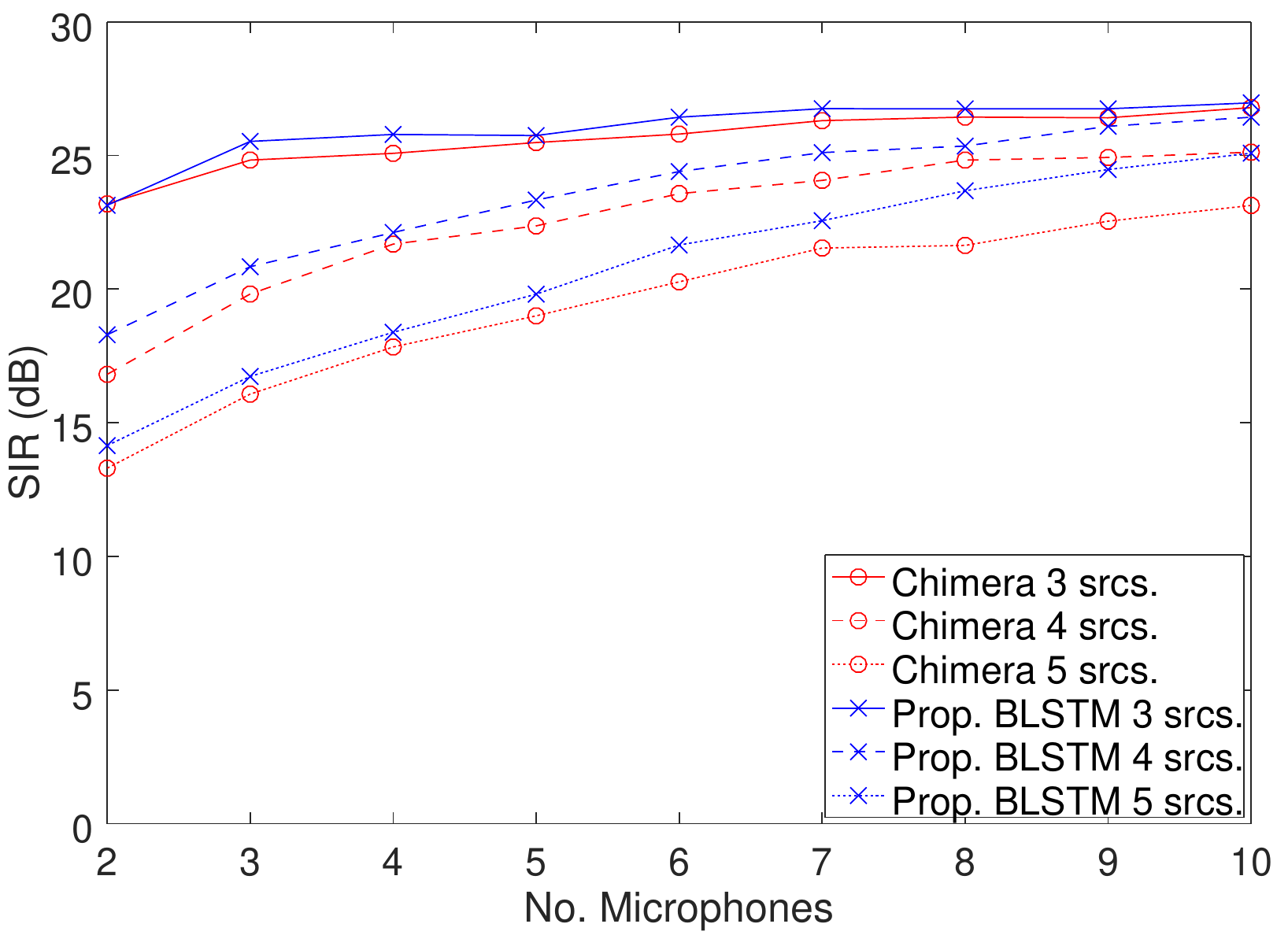}
\caption{Number of microphones vs SIR of the output of the trained models.}
\label{fig:SIRvsMIC}
\end{figure}

In Figure \ref{fig:SIRvsMIC}, the SIR performance is shown for both the recommended configuration of the BLSTM model as well as the best performing configuration of the Chimera when the number of microphones of the linear array is increased up to 10 microphones. No re-training was carried out and the same sources were used throughout the increase in number of microphones. 

\begin{figure}[ht]
\centering
\includegraphics[width=0.48\textwidth]{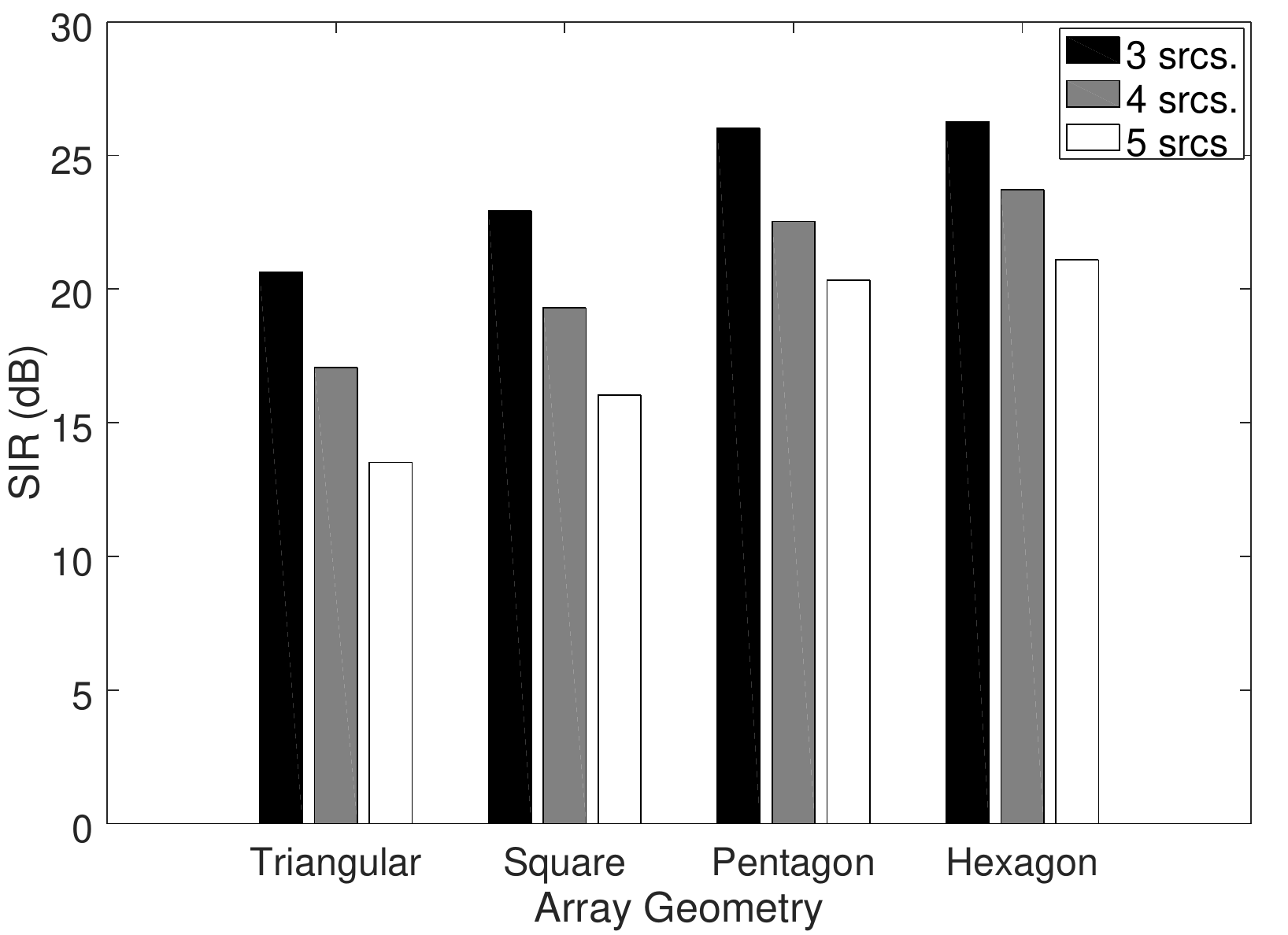}
\caption{Array geometry vs SIR of the output of the recommended BLSTM architecture configuration.}
\label{fig:SIRvsGEOM}
\end{figure}

Additionally, to investigate the impact of changing the geometry of the simulated microphone array, the SIR performance as the number of sources increases when using a linear, triangle, square, pentagonal and hexagonal array is shown in Figure \ref{fig:SIRvsGEOM}.

\begin{figure}[ht]
\centering
\includegraphics[width=0.48\textwidth]{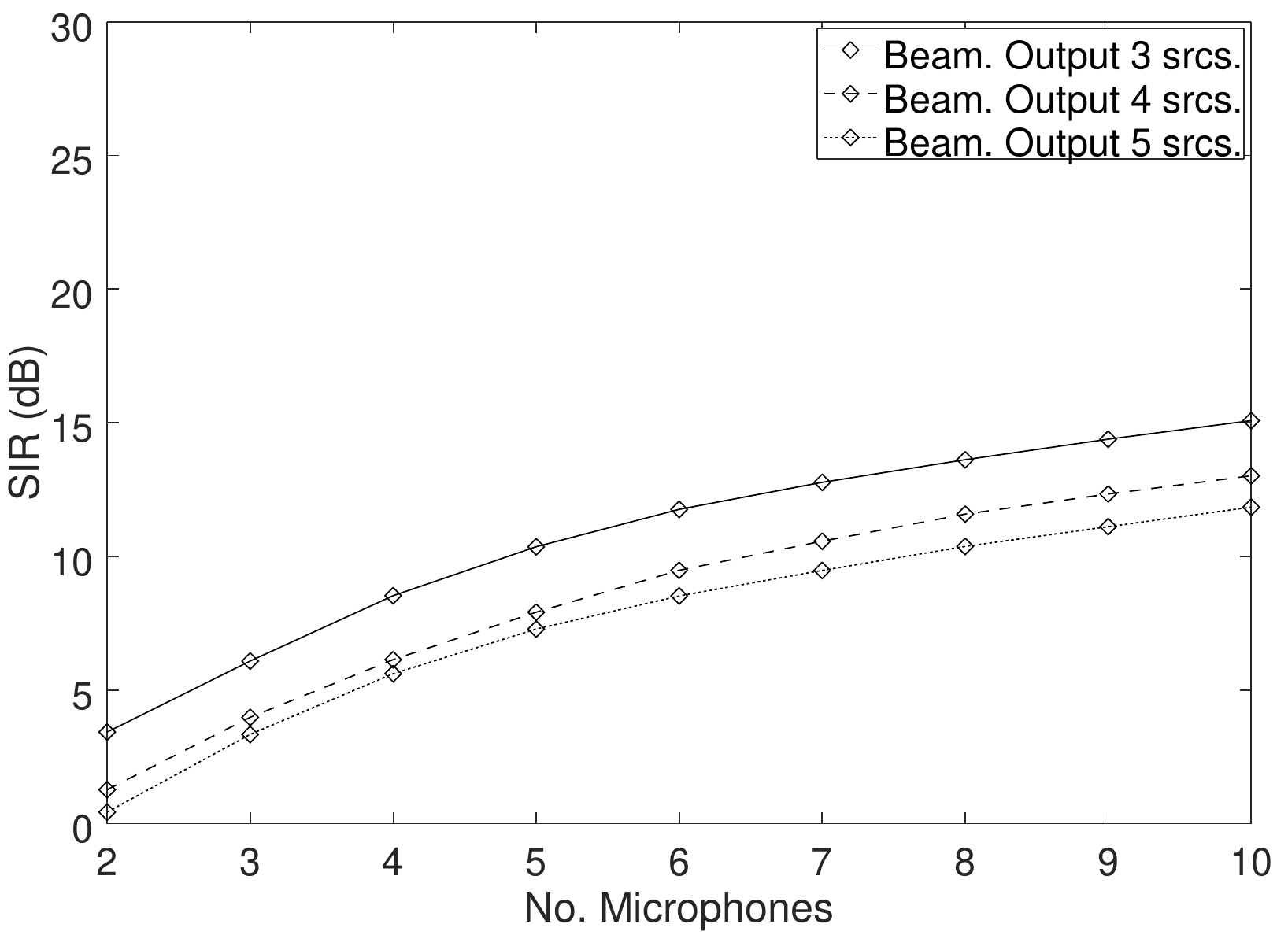}
\caption{Number of microphones vs SIR of the beamformer output.}
\label{fig:SIRvsMICBeam}
\end{figure}

In both Figures \ref{fig:SIRvsMIC} and \ref{fig:SIRvsGEOM}, the same tendency observed in the previous section is still present: as the number of sources increases, the SIR decreases. More on topic, it can also be seen that, overall, as the number of microphones increases, so does the SIR. A possible explanation for this is that the quality of the beamformer output is affected by the number of microphones used, as shown in Figure \ref{fig:SIRvsMICBeam}.

When comparing the SIR of the beamformer output (reported in Figure \ref{fig:SIRvsMICBeam}) and the SIR of the overall system output (reported in Figures \ref{fig:SIRvsMIC} and \ref{fig:SIRvsGEOM}), a substantial SIR increase can be observed in all the tested numbers of microphones, ranging from 10 to 20 dB difference in performance. This indicates that the BLSTM-based TF binary masking stage is essential in obtaining the reported performance.

More importantly, it is clear that the proposed system is quite robust against changes in the microphone geometry (being linear and the tested 2D geometries). In fact, in most cases, the SIR performance increases when more microphones are added, regardless of the employed geometry.

\section{Results Discussion}
\label{sec:discuss}

It is important to point out that the recommended configuration of the proposed BLSTM model not only provides comparable SIR performance to the Chimera model, but in a considerable amount of cases, it actually outperformed it. The reason this is important is that such a configuration only occupies nearly 10\% of the amount of memory that the Chimera model occupies.

Moreover, in a considerable amount of cases both models provided a SIR close to or above the 20 dB mark, which can be considered as a high level of SIR for most auditory scene analysis~\cite{7179045}.

Additionally, it can be seen in Table \ref{architecture_results_3} that the proposed architectures configured with $L = 3$ obtain a higher SIR than their counterparts with $L = 1$ and $L = 5$, while keeping every other parameter the same. A possible explanation is that this number of BLSTM stacked layers may be a kind of ``sweet spot'' in the established solution space. However, this definitely merits further investigation.

It is also important to mention that the response time of all of the proposed architecture configurations is smaller than the length of the time window that it is fed. Meaning, all these architectures are able to carry out online sound source separation (although with up to a 1-second delay; 0.5-second delay, if using the other recommended configuration in italics in Table \ref{architecture_results_3}). Is is also worth considering that the computer used for these evaluations has an i7-4700MQ at 2.4 GHz (which is a moderate CPU by today's standards), and no GPU was used to run the evaluated configurations. This means that the proposed system provides a high separation performance (an average SIR higher than 20 dB), with moderate computational requirements.

\section{Conclusion}
\label{sec:conclusion}

There is a growing interest in online sound source separation in several areas of application. Deep learning techniques have reached an important level of performance, but require considerable computational resources. In this work, we propose a two step system that first carries out a preliminary estimation of both the source of interest and the cumulative environmental interference, via phase-based frequency masking. These two estimations are then fed to a BLSTM-based model that aims to estimate a time-frequency binary mask that, when applied to the signal of the reference microphone, provides a separation of the source of interest from the cumulative environmental interference.

The system was compared to a variation of the Chimera model, which applies deep clustering to solve the permutation problem encountered when carrying out sound source separation. It was shown that the proposed BLSTM-based system achieved comparable results and even in some cases even obtained slightly higher SIR results. And, it accomplished this only using nearly 10\% of the memory occupied by the Chimera model in a moderately equipped computer. The reasoning behind this is that the first stage of the system (the phase-based beamformer) is solving the permutation problem from the beginning and, thus, the deep clustering parts of the Chimera model are not necessary to properly separate the source of interest.

The results shown here were all carried out with simulated data, with no noise and reverberation present. To this effect, for future work, we propose to investigate several methods of data augmentation that adds this effects to the data, to achieve acceptable SIR performance in real-life scenarios. We also propose to employ the AIRA corpus~\cite{Rascon:2018:AIRA} to evaluate this next version of the proposed system. And finally we will reduce the 1-second delay the system presents by a combination of low-grade GPUs (that still keep the computational requirements low) and shifting processing buffers.

\section*{Acknowledgements}
This work was supported by CONACYT through the project 251319 and PAPIIT. The authors would also like to thank David Kant, from UC Santa Cruz, for his insight during the initial development of the phase-based beamformer here described.

\small{
	\bibliographystyle{cys}
	\bibliography{references}
}
\normalsize

\begin{biography}[]{} 
\end{biography}

{\vskip 12pt}
\noindent
\footnotesize {\textit{Article received on XX/YY/2020; accepted on XX/YY/2020.\\
		Corresponding author is Caleb Rascon.}

\end{document}